\documentclass[11pt]{article}

\usepackage[utf8]{inputenc}
\usepackage[T1]{fontenc}
\usepackage{amsmath,amssymb,bm}
\usepackage{array}
\usepackage{booktabs}
\usepackage{geometry}
\usepackage{graphicx}
\usepackage{microtype}
\usepackage{xcolor}
\usepackage{hyperref}
\usepackage{authblk}
\usepackage{caption}
\usepackage{float}
\usepackage{placeins}
\usepackage{algorithmic}
\usepackage{tikz}
\usepackage{pgfplots}
\usepackage{enumitem}

\floatstyle{ruled}
\newfloat{algorithm}{tb}{loa}
\floatname{algorithm}{Algorithm}

\usetikzlibrary{arrows.meta,positioning,fit,backgrounds}
\pgfplotsset{compat=1.18}

\hypersetup{colorlinks=true,linkcolor=blue!60!black,citecolor=blue!60!black,urlcolor=blue!60!black}
\setlist{nosep,leftmargin=*}


\definecolor{navy}{HTML}{17365D}
\definecolor{bluefill}{HTML}{D9E8FB}
\definecolor{teal}{HTML}{087F8C}
\definecolor{tealfill}{HTML}{D8F0F0}
\definecolor{amber}{HTML}{A96500}
\definecolor{amberfill}{HTML}{FCE8C3}
\definecolor{grayfill}{HTML}{ECEFF3}
\definecolor{redsoft}{HTML}{A33A3A}

\title{\textbf{Grouped Value Attention: Efficient KV Caching via On-Demand Key Reconstruction}}
\author[ ]{Vishesh Tripathi\textsuperscript{*} \quad
Abhay Kumar\textsuperscript{*} \quad
Ramsha Khan\textsuperscript{\ensuremath{\dagger}}}
\affil[ ]{FrontiersMind}
\date{}

\newcommand{\softmax}{\operatorname{softmax}}

\begin{document}
\maketitle
\begingroup
\renewcommand{\thefootnote}{\fnsymbol{footnote}}
\footnotetext[1]{First authors: Vishesh Tripathi and Abhay Kumar.}
\footnotetext[2]{Second author: Ramsha Khan.}
\footnotetext[0]{\texttt{\{vishesh.tripathi,a.kumar,ramsha.khan\}@frontiersmind.ai}}
\endgroup

\begin{abstract}
The KV cache is a primary bottleneck for Transformer decoding: its memory footprint and cache-read traffic grow with sequence length. Grouped-query attention (GQA) reduces this cost by sharing key--value heads, but still stores both a key and a value at every step. We introduce Grouped Value Attention (GVA), which stores grouped values and reconstructs content keys with a learned linear map. At inference, the map can be absorbed into the query, eliminating the need to materialize content keys in the intended decode path. A small shared decoupled RoPE channel retains positional information through a separately cached positional key. For the configurations studied, this representation reduces persistent cache scalars by approximately 45--47\% relative to matched GQA. At the 350M-parameter scale with 30B FineWeb-Edu tokens, the 16-dimensional positional variant reaches 44.35 average accuracy across five tasks, compared with 44.36 for GQA and 43.88 for MLA. These results demonstrate near-GQA benchmark accuracy with a more compact cache representation. To translate this compact representation into faster autoregressive inference, we have developed custom decoding kernels and are currently evaluating their end-to-end inference performance with an open-source release planned soon.
\end{abstract}

\section{Introduction}

The Transformer computes attention from query, key, and value representations~\cite{vaswani2017attention}.
In autoregressive decoding only the newest query is required,
while the keys and values of all preceding tokens are reused. Implementations
therefore keep a KV cache. Its size grows linearly with context length and often
becomes a dominant capacity and bandwidth cost in long-context serving~\cite{shazeer2019mqa,kwon2023pagedattention}.

Grouped-query attention (GQA) reduces this cost by sharing key--value heads across
groups of query heads~\cite{ainslie2023gqa}. GQA still writes both a key and a value at
every step, so the cache remains two streams. Multi-head latent attention (MLA)
compresses keys and values into a joint latent~\cite{deepseek2024v2}. That shrinks the
cache further, at the price of an extra projection and a more involved decode path.

We introduce Grouped Value Attention (GVA). GVA keeps GQA's grouping on the value but
not on the key: it caches $G$ grouped value streams and reconstructs a distinct
content key for each of the $H$ query heads through a learned per-head linear map,
\begin{equation}
  K_h = V_{g(h)} M_h, \qquad h = 1,\dots,H,
  \label{eq:intro-gva}
\end{equation}
where $g(h)$ is the value group assigned to query head $h$. Values already carry the
content delivered to the attention output; $M_h$ selects the features head $h$ uses
for scoring. There is no separate key projection, and no content key is written to
the cache. Where the head and group indices are not needed we abbreviate
\eqref{eq:intro-gva} as $K = VM$. Because each $M_h$ is fixed at inference, it can be
absorbed into the query: the content score against each cached position becomes a
single inner product with the stored value, so the intended decode path need not
write a content-key stream.

Standard RoPE applied to the query and the reconstructed key breaks this absorption.
The rotation between a query and a cached key depends on the position \emph{pair}, so
it cannot be folded into a single transformed query reused across cached positions;
retaining it would require either position-dependent key reconstruction or a full key
cache. We therefore adopt a small shared decoupled RoPE channel, following DeepSeek
MLA~\cite{deepseek2024v2}: an unrotated content slice reconstructed from the stored value,
plus a short rotated slice shared across heads. Position is retained without restoring
a full key cache, and because the positional key is shared it adds only $d_r$ scalars
per token rather than $G d_r$.

The systems case is direct. Token-by-token decoding repeatedly reads the attention
cache and is often limited by memory bandwidth. GVA targets faster, more
memory-efficient autoregressive inference by reducing the intended persistent cache by
approximately 45--47\% relative to matched GQA. Prefill is compute-bound rather than
cache-bound, so GVA seeks no advantage there; its prefill cost stays close to GQA
provided the combined content and positional width does not exceed the head-dimension
tile the baseline already occupies (Section~\ref{sec:limitations}). Neither latency nor
decode-throughput gains are established by the present experiments. Compared with MLA,
GVA uses the value itself as the persistent content state rather than a separate joint
latent, preserving a direct value path.

\begin{figure}[!b]
\centering
\includegraphics[width=\textwidth]{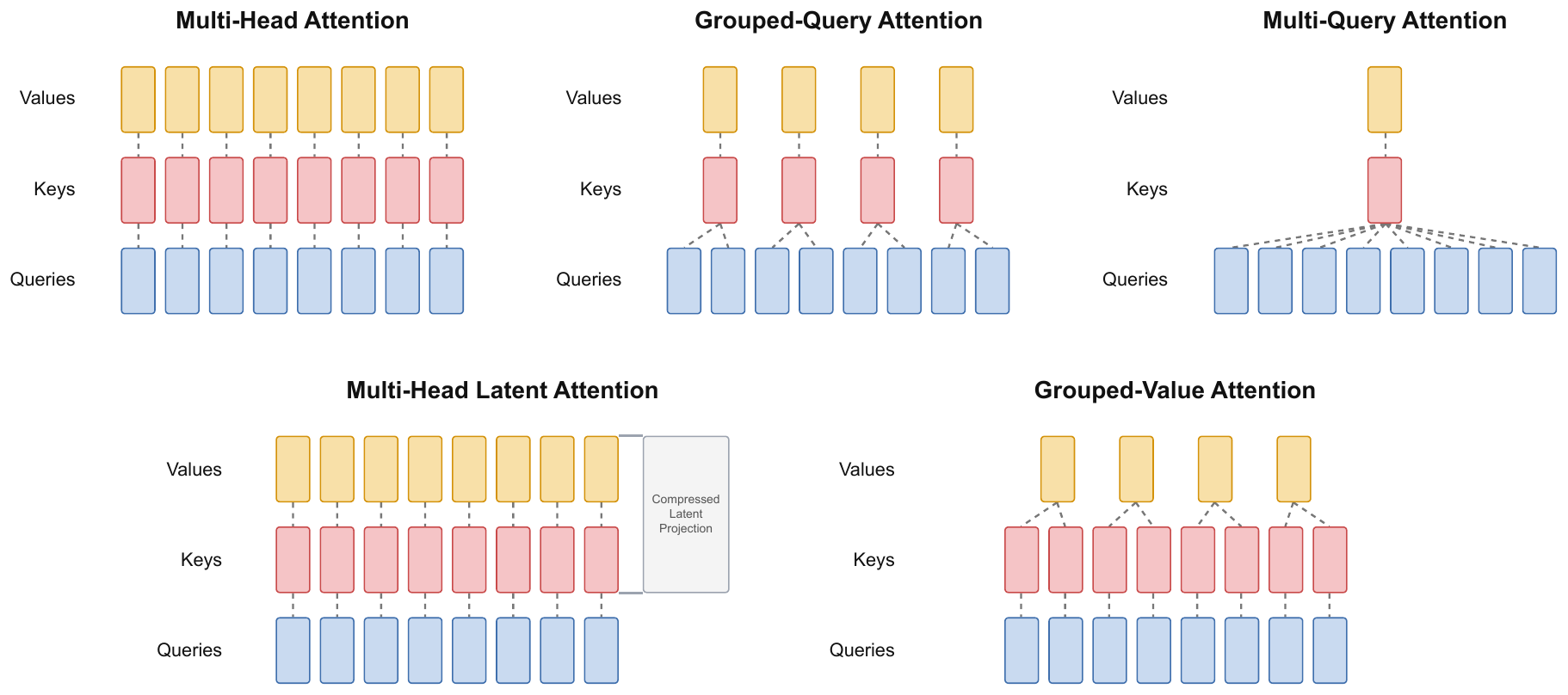}
\caption{Attention caching strategies for eight query heads. GVA caches four grouped values and reconstructs content keys as $K=VM$. The depicted content cache is half the size of matched GQA; including the shared positional key, the total intended cache is approximately 45--47\% smaller for the configurations studied.}
\label{fig:overview-of-attention-variants.png}
\end{figure}

Figure~\ref{fig:overview-of-attention-variants.png} provides an overview of the attention caching strategies and
GVA's value-only content cache.

Our contributions are:
\begin{itemize}
  \item Grouped Value Attention, which caches grouped values and reconstructs per-head
        content keys with a linear map that is absorbed into the query at inference,
        removing the content-key stream from the cache.
  \item An asymmetry between key and value grouping: GVA reconstructs $H$ distinct
        content keys from only $G$ cached value streams, whereas GQA gives every query
        head in a group the same key. Head-specific content keys are retained while the
        persistent cache shrinks.
  \item A small shared decoupled RoPE channel that keeps positional encoding compatible
        with this reconstruction, at $d_r$ extra cached scalars per token.
  \item The proposed decoupled-RoPE GVA representation reduces cache scalars by
        approximately 45--47\%; its 16-dimensional positional variant reaches 44.35
        average accuracy against 44.36 for GQA---a gap of 0.01 points, within seed
        variation---and 43.88 for MLA, with scores averaged over three runs using
        different random seeds.
\end{itemize}

\section{Related Work}
\label{sec:related-work}

\paragraph{Notation.}
Let \(H\) be the number of query heads, \(G\) the number of key--value groups,
\(d_h\) the head width, and \(T\) the cached length. Per layer and sequence,
multi-head attention stores separate keys and values for each head~\cite{vaswani2017attention,shazeer2019mqa}:
\begin{equation}
  N_{\mathrm{MHA}} = 2\,T H d_h.
\end{equation}

\paragraph{Multi-query attention.}
Autoregressive decoding is limited by the bandwidth of loading keys and values
at every step~\cite{shazeer2019mqa}. Multi-query attention (MQA) keeps \(H\)
query heads but a single shared key--value head (\(G{=}1\))~\cite{shazeer2019mqa}. The cache is
\begin{equation}
  N_{\mathrm{MQA}} = 2\,T d_h,
\end{equation}
a factor of \(H\) smaller than multi-head attention. That speeds up decoding,
at the cost of quality and, in some settings, training
stability~\cite{shazeer2019mqa,ainslie2023gqa}.

\paragraph{Grouped-query attention.}
GQA interpolates between multi-head attention and MQA: query heads are split
into \(G\) groups, and each group shares one key--value
head~\cite{ainslie2023gqa}. The cache is
\begin{equation}
  N_{\mathrm{GQA}} = 2\,T G d_h.
\end{equation}
The cases \(G{=}H\) and \(G{=}1\) recover multi-head attention and MQA~\cite{ainslie2023gqa}. With a
modest \(G\), GQA achieves quality close to multi-head attention and
speed close to MQA in the experiments of Ainslie et al.~\cite{ainslie2023gqa}.
It is also used in open models such as Llama 2 70B~\cite{touvron2023llama2}.

\paragraph{Multi-head latent attention.}
DeepSeek MLA compresses keys and values into a low-rank latent of width
\(d_c\), plus a small shared RoPE slice of width \(d_r\), and caches that
instead of full heads~\cite{deepseek2024v2}:
\begin{equation}
  N_{\mathrm{MLA}} = T(d_c + d_r).
\end{equation}
At decode time the key up-projection can be absorbed into the query projection so that attention
need not materialize a full content-key tensor~\cite{deepseek2024v2}.

\section{Grouped Value Attention}

We hypothesize that the content key can be derived from the value, as both
encode information about the same underlying content, eliminating the need
for a separate content-key cache.

GVA is built on GQA grouping: \(H\) query heads share \(G\) value heads.
Only the key changes. We first reused the value as the key, then replaced
that with a linear map from the stored value. We use row vectors throughout
this section; \(g(h)\) denotes the value group assigned to query head \(h\).

\subsection{Value-only cache and \texorpdfstring{\(K_h = V_{g(h)} M_h\)}{K(h) = V(g(h)) M(h)}}
\label{sec:gva-kvm}

GQA caches both a grouped key and a grouped value. The two streams have the
same shape, so the simplest cut is to store one of them. Values must persist,
because they are what attention aggregates.

\paragraph{Shared KV.}
Our first design dropped the key projection and set \(K = V\). The cache is
then exactly half of GQA, but the training loss never recovered to the GQA
baseline (Figure~\ref{fig:shared-kv}). One
vector is asked both to score and to be retrieved.
We trained Lumma-0.6B using this initial shared-KV approach with query
normalization, which yielded a slight improvement over shared KV without
query normalization, and open-sourced it on Hugging Face as
Lumma-0.6B-Base~\cite{frontiersmind2026lumma}.

\paragraph{Linear reconstruction.}
We therefore keep a dedicated value and reconstruct a content key for each
query head,
\begin{equation}
  \label{eq:kvm}
  K_h = V_{g(h)} M_h,
  \qquad
  M_h \in \mathbb{R}^{d_h \times d_n},
  \qquad h = 1,\ldots,H.
\end{equation}
Here \(d_n\) is the content-key width, with \(d_n=d_h\) when there is no
positional slice. GVA uses one reconstruction map \(M_h\) per query head.
We write \(M\) below, suppressing the head and group indices when discussing
a single head.
There is no independent key projection. Only the grouped values
\(V_{g(h)}\) are written to the content cache. Values carry the content
delivered to the output; \(M_h\) selects the features used for scoring.
The per-head maps retain head-specific content keys at a small,
sequence-independent parameter cost.

\paragraph{Key diversity.}
With per-head maps, GVA caches \(G\) value streams but can produce \(H\)
distinct content-key streams through reconstruction.
In GQA, every query head within a group scores against the same key vector;
GVA instead removes this key sharing and recovers head-specific key
diversity with a smaller persistent cache than GQA for the configurations
studied.
These keys remain linear transforms of their grouped values, rather than
unconstrained independent projections.
This is an advantage over GQA, not MLA: MLA also uses per-head key
up-projections over a shared latent~\cite{deepseek2024v2}.

Table~\ref{tab:key-diversity} summarizes the content-key and cache-stream counts.
\begin{table}[!htbp]
\centering
\caption{Head-specific content-key representations and persistent
cache streams per layer. The shared positional stream has width \(d_r\).}
\label{tab:key-diversity}
\begin{tabular}{lcc}
\toprule
Method & Distinct content keys & Cached streams \\
\midrule
MHA & \(H\) & \(2H\) \\
GQA & \(G\) & \(2G\) \\
GVA & \(H\) & \(G\) (plus shared \(k^{\mathrm{rope}}\)) \\
\bottomrule
\end{tabular}
\end{table}

\paragraph{Initial scale.}
GQA obtains \(Q\) and \(K\) from separate projections of the same hidden
state, so a common initialization convention provides a reference for
both scales rather than guaranteeing that they match.
In GVA, \(K_h=V_{g(h)}M_h\) is a projection of a projection; in our initial
runs, default initialization of \(M_h\) left keys well below queries in
scale, producing nearly uniform attention and spending early training
recovering from this mismatch.
We therefore initialize each map to match the initial RMS of content keys
and queries using \(\sigma_M=\sigma_Q/(\sigma_V\sqrt{d_{\mathrm{in}}})\),
where \(d_{\mathrm{in}}=d_h\) is the value width contracted over by the map
and \(\sigma_Q\) is measured after any query normalization.
Appendix~\ref{app:scale} gives the derivation and its assumptions.

\subsection{Absorption at decode}
\label{sec:gva-absorb}

Because \(M_h\) is fixed at inference, the content key never has to be materialized.
For query head \(h\), the content score against a cached position \(j\) is
\begin{equation}
  \label{eq:absorb}
  q_h k_{j,h}^\top
  = q_h (v_{j,g(h)} M_h)^\top
  = (q_h M_h^\top)v_{j,g(h)}^\top
  = \tilde{q}_h v_{j,g(h)}^\top,
\end{equation}
where \(\tilde{q}_h = q_h M_h^\top\) is computed once per query token and
head. Here \(q_h\) and \(k_{j,h}\) denote the content slices when a separate
positional slice is present. We suppress the head and group indices in the
single-head expressions below. Content attention then reads only the value cache. This identity is exact. In the fused decode
formulation, attention uses \(\tilde{q}\) directly over the value cache,
avoiding storage of a temporary key tensor.

Standard RoPE on the query and reconstructed key breaks the absorption
in~\eqref{eq:absorb}. With head and group indices suppressed, let \(q_t\)
denote the unrotated query at position \(t\). In row-vector notation,
\[
  q_t^{\mathrm{rot}} = q_t R_t^\top,
  \qquad
  k_j^{\mathrm{rot}} = (v_j M)R_j^\top,
\]
so the score is
\[
  q_t^{\mathrm{rot}}(k_j^{\mathrm{rot}})^\top
  = q_t R_t^\top R_j M^\top v_j^\top
  = q_t R_{j-t} M^\top v_j^\top,
  \qquad R_{j-t}=R_t^\top R_j.
\]
The relative rotation depends on the query--key position pair and sits
between \(q_t\) and \(M^\top\). Even for a fixed query position \(t\), it
varies with the cached position \(j\), so for a general learned \(M\) it
cannot be folded into a single transformed query reused across all cached
positions.

\subsection{Decoupled RoPE}
\label{sec:gva-drope}

We split each query and key head into an unrotated content slice of width \(d_n\)
and a rotated positional slice of width \(d_r\), while values retain width \(d_h\).
We adopt the decoupled RoPE strategy of DeepSeek MLA~\cite{deepseek2024v2}:
RoPE is applied only to the positional slice, and the positional key is
shared across heads:
\begin{align}
  k^{\mathrm{nope}}_{j,h} &= v_{j,g(h)} M_h,
  \qquad
  M_h \in \mathbb{R}^{d_h \times d_n},
  \label{eq:knope} \\
  k^{\mathrm{rope}}_j &= (x_j W_r)R_j^\top,
  \qquad
  W_r \in \mathbb{R}^{d_{\mathrm{model}} \times d_r}.
  \label{eq:krope}
\end{align}
The query is split the same way, with a per-head positional slice rotated
by \(R_t\) at query position \(t\). Suppressing head and group indices, the score is
\begin{equation}
  q k_j^\top
  =
  q^{\mathrm{nope}}(k^{\mathrm{nope}}_j)^\top
  +
  q^{\mathrm{rope}}(k^{\mathrm{rope}}_j)^\top.
\end{equation}
The first term still absorbs:
\(q^{\mathrm{nope}}(v_j M)^\top = (q^{\mathrm{nope}}M^\top)v_j^\top\).
The second term uses the cached, already-rotated \(k^{\mathrm{rope}}_j\).
Nothing learned sits between the two rotations, so relative position is
preserved. Because \(k^{\mathrm{rope}}\) is shared, it costs \(d_r\)
scalars per token, not \(G d_r\). The full query/key width used for score
scaling is \(d_{\mathrm{qk}}=d_n+d_r\).

\subsection{Cache size}
\label{sec:gva-cache}

GQA stores two grouped streams,
\begin{equation}
  N_{\mathrm{GQA}} = 2\,T G d_h.
\end{equation}
Shared KV and GVA without a positional slice store only values,
\(T G d_h\), exactly half. With decoupled RoPE the persistent state is
\begin{equation}
  \label{eq:ngva}
  N_{\mathrm{GVA}} = T G d_h + T d_r,
\end{equation}
so
\begin{equation}
  \frac{N_{\mathrm{GVA}}}{N_{\mathrm{GQA}}}
  = \frac{1}{2} + \frac{d_r}{2 G d_h}.
\end{equation}
The second term is a few percent for the widths we use, which is why we say
GVA \emph{roughly} halves the GQA cache. Adding positional dimensions on top of the
content width changes parameters and the query/key width, not the form of~\eqref{eq:ngva}:
\(V\) stays \(d_h\)-wide and \(k^{\mathrm{rope}}\) stays shared; any change in
\(d_r\) is accounted for by the \(T d_r\) term.

\begin{algorithm}[t]
\caption{One decode step of GVA with decoupled RoPE}
\label{alg:gva-decode}
\small
\begin{algorithmic}[1]
\REQUIRE Current hidden state \(x_t\); maps \(W_Q, W_V, W_r, \{M_h\}_{h=1}^{H}\);
caches \(V_{1:t-1,g}\) for \(g=1,\ldots,G\) and \(k^{\mathrm{rope}}_{1:t-1}\).
\ENSURE Per-head attention outputs \(\{o_{t,h}\}_{h=1}^{H}\).
\STATE \(q_t \leftarrow x_t W_Q\); split each head into \(q^{\mathrm{rope}}_{t,h},\, q^{\mathrm{nope}}_{t,h}\)
\STATE \(v_t \leftarrow x_t W_V\)
\FOR{each value group \(g=1,\ldots,G\)}
  \STATE \(V_{1:t,g} \leftarrow [V_{1:t-1,g}; v_{t,g}]\)
\ENDFOR
\STATE \(k^{\mathrm{rope}}_t \leftarrow (x_t W_r)R_t^\top\)
\STATE \(k^{\mathrm{rope}}_{1:t} \leftarrow [k^{\mathrm{rope}}_{1:t-1}; k^{\mathrm{rope}}_t]\)
\FOR{each query head \(h=1,\ldots,H\)}
  \STATE \(q^{\mathrm{rope}}_{t,h} \leftarrow q^{\mathrm{rope}}_{t,h}R_t^\top\)
  \STATE \(\tilde{q}_{t,h} \leftarrow q^{\mathrm{nope}}_{t,h}M_h^\top\)
         \COMMENT{absorb; do not form \(K^{\mathrm{nope}}\)}
  \STATE \(s_{t,h} \leftarrow \tilde{q}_{t,h} V_{1:t,g(h)}^\top
         + q^{\mathrm{rope}}_{t,h}(k^{\mathrm{rope}}_{1:t})^\top\)
  \STATE \(o_{t,h} \leftarrow \softmax(s_{t,h}/\sqrt{d_{\mathrm{qk}}})V_{1:t,g(h)}\)
\ENDFOR
\STATE \textbf{return} \(\{o_{t,h}\}_{h=1}^{H}\)
\end{algorithmic}
\end{algorithm}

The usual concatenation and output projection combine the per-head outputs.
Training uses the same content--position split with standard causal attention.
Algorithm~\ref{alg:gva-decode} describes the intended decode path, and the
cache count in~\eqref{eq:ngva} describes its persistent state rather than
measured peak serving memory. We have developed custom decoding kernels and
are currently testing their inference performance, with an open-source
release planned soon.

\FloatBarrier
\section{Experiments}
\label{sec:experiments}

We train decoder-only Transformers from scratch at the 350M-parameter scale
on a 30B-token sample of FineWeb-Edu~\cite{penedo2024fineweb}. All models
use the same data order, token budget, optimizer, and context length unless
a variant is named below. The GQA baseline uses grouped key--value heads;
MLA and every GVA run keep that query-head count so the comparison is on
the cache representation, not on the width of the query.

During pre-training, we used ZClip~\cite{kumar2025zclipadaptivespikemitigation} to mitigate gradient spikes and help prevent loss spikes. ZClip adaptively clips gradients using z-score-based anomaly detection on gradient norms.

\subsection{Variants}
\label{sec:exp-variants}

\paragraph{Shared KV.}
The first cache cut sets \(K = V\) and stores only the grouped value. We
report it as a reference: the cache is exactly half of GQA, but the loss
does not recover (Figure~\ref{fig:shared-kv}). It is not a proposed system.

\paragraph{GQA and MLA.}
GQA caches grouped keys and values~\cite{ainslie2023gqa}. MLA caches a
joint latent plus a shared RoPE slice~\cite{deepseek2024v2}. Both are trained
with the same recipe as GVA.

\paragraph{GVA.}
GVA stores grouped values and reconstructs keys with \(K_h = V_{g(h)}M_h\),
using one reconstruction map per query head in all GVA variants.
We compare:
\begin{itemize}
  \item \textbf{GVA baseline:} linear reconstruction with a standard init
    of \(M\).
  \item \textbf{GVA, scale-matched:} \(M\) initialized so \(K\) and \(Q\)
    start at the same RMS (Appendix~\ref{app:scale}), with query RMSNorm.
  \item \textbf{GVA, variance-fixed:} the same init of \(M\), without
    query RMSNorm. This is our default GVA without decoupled RoPE.
  \item \textbf{GVA + decoupled RoPE:} the default GVA plus a shared
    positional slice of width \(d_r \in \{16, 24\}\)
    (Section~\ref{sec:gva-drope}).
\end{itemize}
Shared KV and GVA without decoupled RoPE apply ordinary RoPE to the
reconstructed key. That path cannot absorb \(M\) at decode; we still
report it because it isolates the reconstruction from the positional
design.

\subsection{Evaluation}
\label{sec:exp-eval}

We report language-model training loss over training steps, and zero-shot
accuracy on HellaSwag~\cite{zellers2019hellaswag},
WinoGrande~\cite{sakaguchi2020winogrande},
OpenBookQA~\cite{mihaylov2018openbookqa}, and the Easy and Challenge splits of
ARC~\cite{clark2018arc}. For each configuration, we perform three runs,
each using a different random seed, and report the arithmetic mean of
the three accuracies for each task. Average is the unweighted mean of
these five task-level means. Small differences should not be interpreted
as statistically significant without assessing variability across seeds.

Cache sizes follow Section~\ref{sec:gva-cache}: GQA stores \(2TGd_h\);
shared KV and GVA without a rope slice store \(TGd_h\); GVA with
decoupled RoPE stores \(TGd_h + Td_r\).

We have developed custom decoding kernels and are currently evaluating
their end-to-end inference performance, including fused decoding throughput,
for comparison with MLA and GQA. These systems measurements are not reported
here; an open-source release is planned soon.

\section{Results}
\label{sec:results}

\begin{figure}[t]
  \centering
  \includegraphics[width=0.9\linewidth]{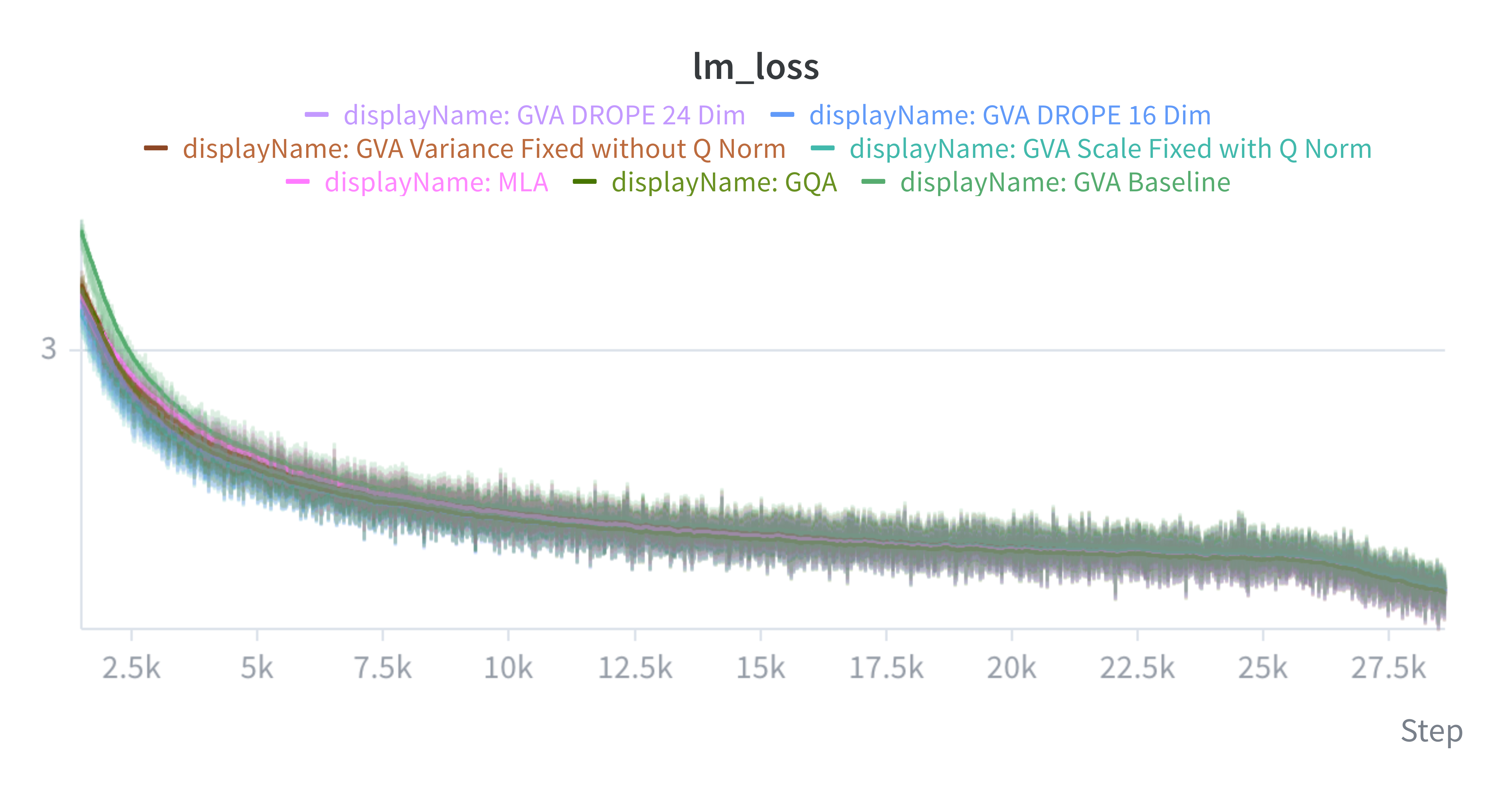}
  \caption{Training loss on FineWeb-Edu for GQA, MLA, and GVA, plotted
  against training steps. After initialization is scale-matched, GVA follows
  a similar trajectory to GQA. Shared KV is omitted here; see
  Figure~\ref{fig:shared-kv}.}
  \label{fig:train-loss}
\end{figure}

\paragraph{Training loss.}
Figure~\ref{fig:train-loss} plots language-model loss for GQA, MLA, and
the GVA variants. After the initial transient, scale-matched GVA tracks GQA
and MLA. The baseline GVA, with standard initialization of \(M\), is worse
early and narrows the gap later, consistent with the scale account in
Appendix~\ref{app:scale}. Decoupled-RoPE runs follow the same broad
trajectory; we do not observe a second collapse once \(M\) is scale-matched.

\begin{figure}[t]
  \centering
  \includegraphics[width=0.9\linewidth]{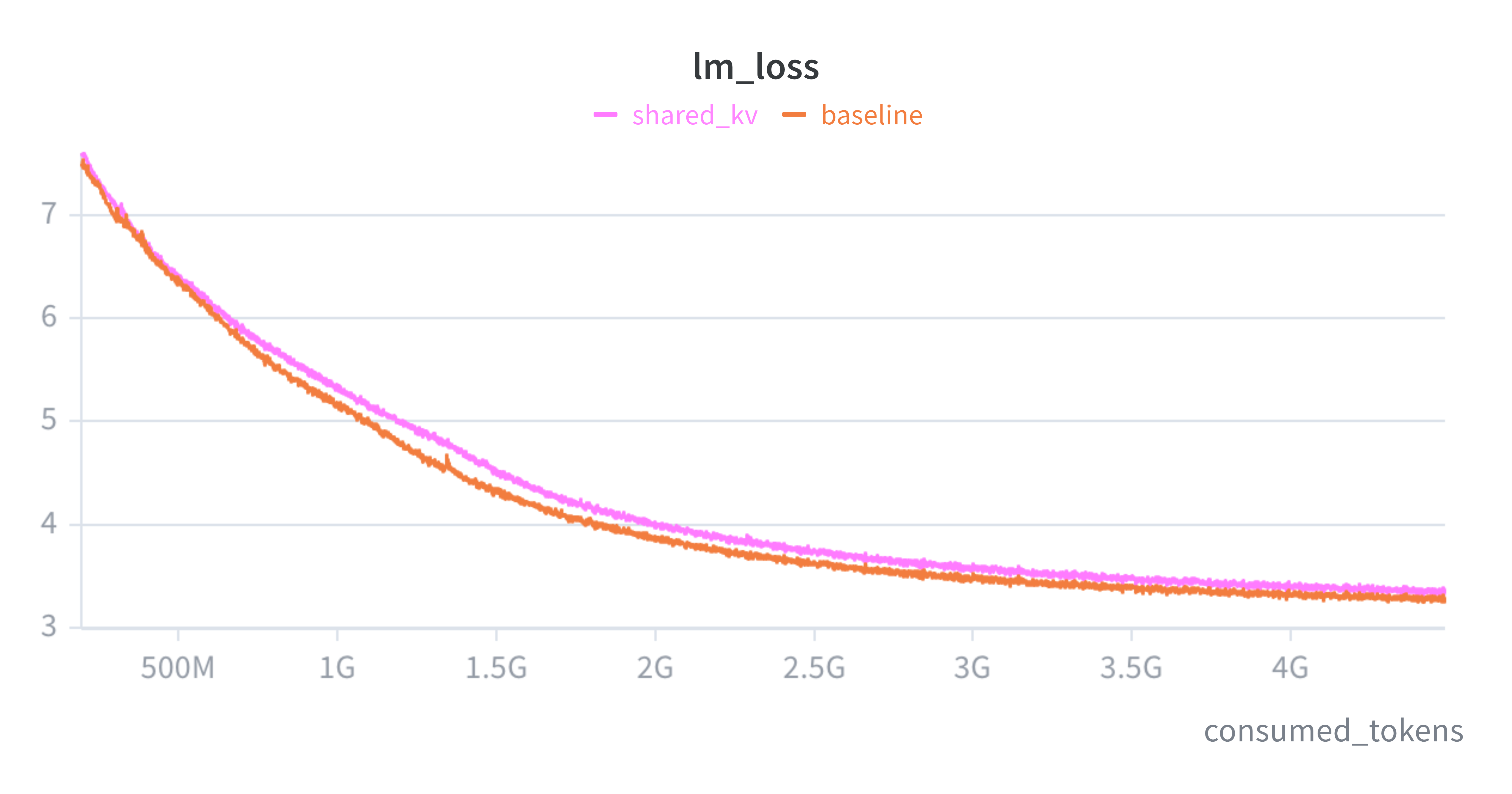}
  \caption{Training loss for shared KV (\(K = V\)) against the GQA
  baseline. The cache is half of GQA, but the gap does not close. This
  motivates reconstructing the key from a dedicated value
  (Section~\ref{sec:gva-kvm}).}
  \label{fig:shared-kv}
\end{figure}

\paragraph{Shared KV.}
Figure~\ref{fig:shared-kv} shows the first cut: \(K = V\) halves the GQA
cache, but the loss stays above GQA for the whole displayed run. A single
vector does not recover the key/value split. All later GVA runs keep a
dedicated value and reconstruct the key.

\paragraph{Downstream accuracy.}
Table~\ref{tab:bench} reports zero-shot accuracy averaged over three
runs with different random seeds for each configuration. Average is the
unweighted mean of the five task-level mean accuracies.

\begin{table}[tb]
\centering
\caption{Accuracy (\%) on five benchmarks, averaged over three runs
with different random seeds for each configuration. Average is the
unweighted mean of the five task-level means. Bold marks the best entry
in each column.}
\label{tab:bench}
\small
\setlength{\tabcolsep}{3pt}
\begin{tabular}{@{}>{\raggedright\arraybackslash}p{4.8cm}rrrrrr@{}}
\toprule
Method & HellaSwag & WinoGrande & OBQA & ARC-E & ARC-C & Avg. \\
\midrule
GQA & \textbf{43.41} & 52.48 & 33.40 & 63.42 & 29.09 & 44.36 \\
MLA & 43.20 & 51.61 & 34.60 & 62.87 & 27.13 & 43.88 \\
GVA baseline & 42.12 & 52.96 & \textbf{34.80} & 61.95 & 27.73 & 43.91 \\
GVA, scale-matched + Q-norm & 42.05 & \textbf{53.51} & 34.40 & 62.75 & \textbf{29.35} & \textbf{44.41} \\
GVA, variance-fixed, no Q-norm & 42.71 & 53.35 & 32.20 & 62.33 & 28.24 & 43.77 \\
GVA + DRoPE \(d_r{=}24\) & 42.94 & 53.12 & 33.20 & 63.69 & 28.50 & 44.29 \\
GVA + DRoPE \(d_r{=}16\) & 42.69 & 53.35 & 33.60 & \textbf{63.81} & 28.32 & 44.35 \\
\bottomrule
\end{tabular}
\end{table}

Scale-matched GVA with query RMSNorm is the strongest GVA row (44.41)
and sits close to GQA (44.36). The default without query RMSNorm is
slightly lower (43.77) but remains close to MLA (43.88). The GVA baseline
without matched initialization reaches 43.91: reconstruction alone is
already in this band, and scale matching primarily improves early loss
rather than consistently improving the final average across variants.

Decoupled RoPE is the proposed serving design. With \(d_r{=}16\), the
average is 44.35, 0.01 percentage points below GQA. With \(d_r{=}24\), it is
44.29. We treat \(d_r{=}16\) as the better operating point among the two
widths tested. These runs do not isolate the effects of positional width
and content-width allocation. Neither DRoPE row beats GQA on the average;
the claim is a much smaller intended cache at near-GQA quality, not a
quality win.

\paragraph{Cache.}
Relative to GQA, shared KV and GVA without a separate positional slice
require half the cache scalars when only values are retained. With DRoPE
the ratio is \(\tfrac12 + d_r/(2Gd_h)\)
(Section~\ref{sec:gva-cache}), corresponding to about 47\% saved at
\(d_r{=}16\) and 45\% at \(d_r{=}24\) for our grouping. These are
representation-level counts, not measured serving-memory reductions.
Ordinary-RoPE variants cannot use the absorbed decode path and require
position-dependent key reconstruction.

\section{Limitations}
\label{sec:limitations}
\label{sec:limits}

The roughly 45--47\% cache reduction for decoupled-RoPE GVA describes the
\emph{intended} decode state: grouped \(V\) plus a shared
\(k^{\mathrm{rope}}\). Omitting the positional slice gives the idealized
50\% value-only reduction. We have developed custom decoding kernels and
are currently testing their inference performance, with an open-source
release planned soon. Fused decode throughput, peak serving memory, and
batch capacity measurements are not reported here.

All reported comparison runs use one scale (approximately 350M parameters),
one data mix (30B FineWeb-Edu tokens), and three different random seeds
per benchmark configuration, with scores averaged across runs. DRoPE with
\(d_r{=}24\) is behind GQA on average; we have not systematically swept
RoPE width, additive versus carved allocation, or longer contexts. Shared
KV is a failed first cut, not a baseline we recommend.

\section{Conclusion}
\label{sec:conclusion}

GVA stores grouped values and reconstructs keys with a linear map
\(K = VM\). Sharing one vector as both key and value halves the cache
but does not match GQA in the observed run. Reconstructing the key and
matching its initial scale to the query yields quality close to GQA.
A small shared decoupled RoPE channel keeps position compatible with
absorbing \(M\) at decode. Relative to GQA, the intended persistent cache
is roughly half, while downstream quality remains in the same broad band
(Table~\ref{tab:bench}). We have developed custom decoding kernels and are
currently testing their end-to-end inference performance, with an open-source
release planned soon. Further work includes completing this evaluation and
a broader sweep of RoPE width, model scale, and random seeds.

\clearpage
\appendix

\section{Query--Key Scale}
\label{app:scale}

\begin{figure}[b]
  \centering
  \includegraphics[width=\linewidth]{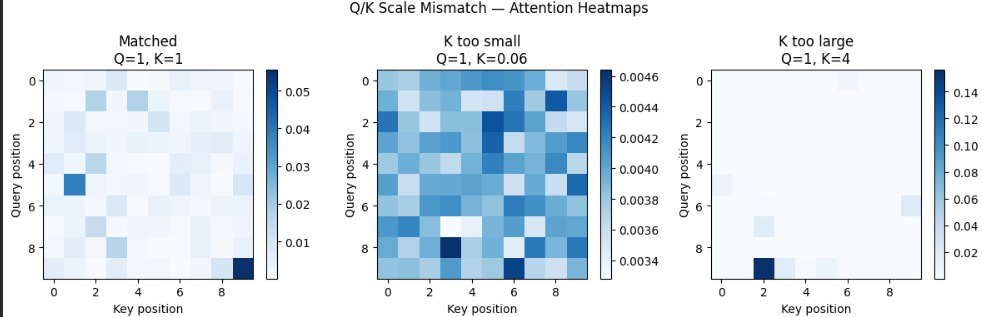}
  \caption{Illustrative attention heatmaps on a 10-position sequence
  with matched query and key scales (left), keys that are too small
  (middle), and keys that are too large (right). Small keys produce
  nearly uniform attention; large keys concentrate the weights.
  Each panel uses its own color scale.}
  \label{fig:scale}
\end{figure}

If query and key coordinates are centered and weakly correlated, with
standard deviations \(\sigma_Q\) and \(\sigma_K\), then for row vectors
\begin{equation}
  \operatorname{Var}\!\left(\frac{q k^\top}{\sqrt{d_h}}\right)
  \approx \sigma_Q^2\sigma_K^2.
  \label{eq:logit-variance}
\end{equation}
When the product is too small, softmax sees nearly equal logits and
attention is flat. When it is too large, a few positions dominate.
Figure~\ref{fig:scale} shows both regimes.

GQA obtains \(Q\) and \(K\) from separate projections of the same hidden
state, providing a reference for their initial scales rather than a
guarantee that they match. In GVA, \(K=VM\), so the scale of \(K\) also
depends on \(M\). In our initial runs, default initialization left \(K\)
far below \(Q\), and early training was spent adjusting this mismatch.

\paragraph{Initialization.}
Let \(d_{\mathrm{in}}\) be the input width of \(M\), namely the value
width. Assuming approximately centered value coordinates with variance
\(\sigma_V^2\), independent of the initial map entries, we draw
\begin{equation}
  M_{ij} \sim \mathcal{N}\!\left(0,\;
  \frac{\sigma_Q^2}{\sigma_V^2\,d_{\mathrm{in}}}\right).
  \label{eq:scale-init}
\end{equation}
Then \(\operatorname{Var}(K_j)\approx
 d_{\mathrm{in}}\sigma_V^2\operatorname{Var}(M_{ij})=\sigma_Q^2\),
so \(K\) starts at approximately the same RMS as \(Q\). The target uses
the query scale after any query normalization. After initialization,
\(M\) remains learnable. Optional query RMSNorm~\cite{zhang2019rmsnorm}
controls the query scale, but the variance-fixed row in
Table~\ref{tab:bench} remains competitive without it. This initialization
does not normalize the cached values.

\paragraph{Decoupled RoPE.}
Apply the same rule per slice. The fan-in of \(M\) is the value width,
not the content width. Initialize the shared positional-key projection
so \(k^{\mathrm{rope}}\) matches the initial RMS of
\(q^{\mathrm{rope}}\). Assess positional and content scales separately:
pooling them into one key variance hid the mismatch in early runs.

\begin{figure}[tbp]
  \centering
  \includegraphics[width=\linewidth]{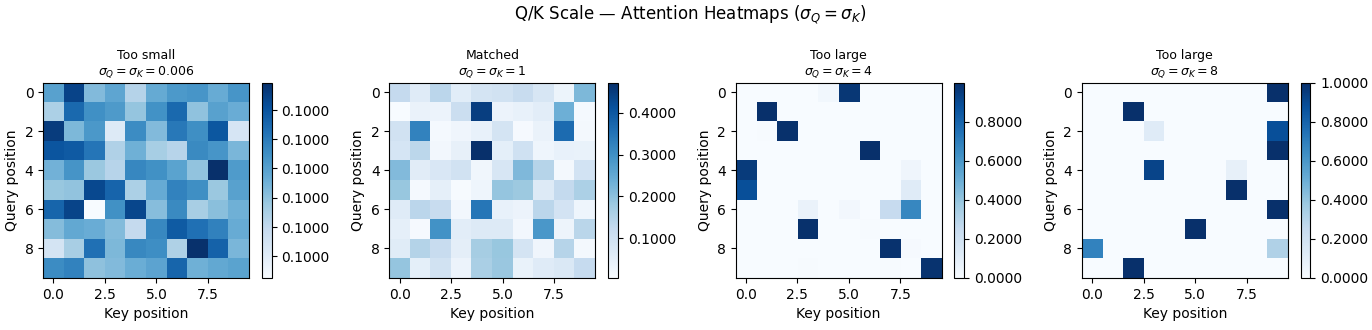}
  \caption{Attention heatmaps ($10\times10$) with matched scales
$\sigma_Q=\sigma_K\in\{0.006,1,4,8\}$ (left to right). Small $\sigma$
gives nearly uniform weights ($\approx 0.1$ per key); unit $\sigma$
gives moderate variation; large $\sigma$ gives peaked, near one-hot
rows. Each panel uses its own color scale. Absolute scale controls
whether softmax is flat, selective, or peaked.}
  \label{fig:joint-qk-scale}
\end{figure}

Figure~\ref{fig:joint-qk-scale} complements Figure~\ref{fig:scale}
by varying query and key scales together, showing that equal scales
alone do not prevent flat or highly concentrated attention.

\clearpage
\section{Extra Evaluation Curves}
\label{app:curves}

Per-task accuracy over training for the seven configurations in
Table~\ref{tab:bench}. The plots retain the original run names and
horizontal-axis units.

\begingroup
\captionsetup{font=small,hypcap=false}
\medskip
\noindent\begin{minipage}{\textwidth}
\centering
\includegraphics[width=0.95\textwidth]{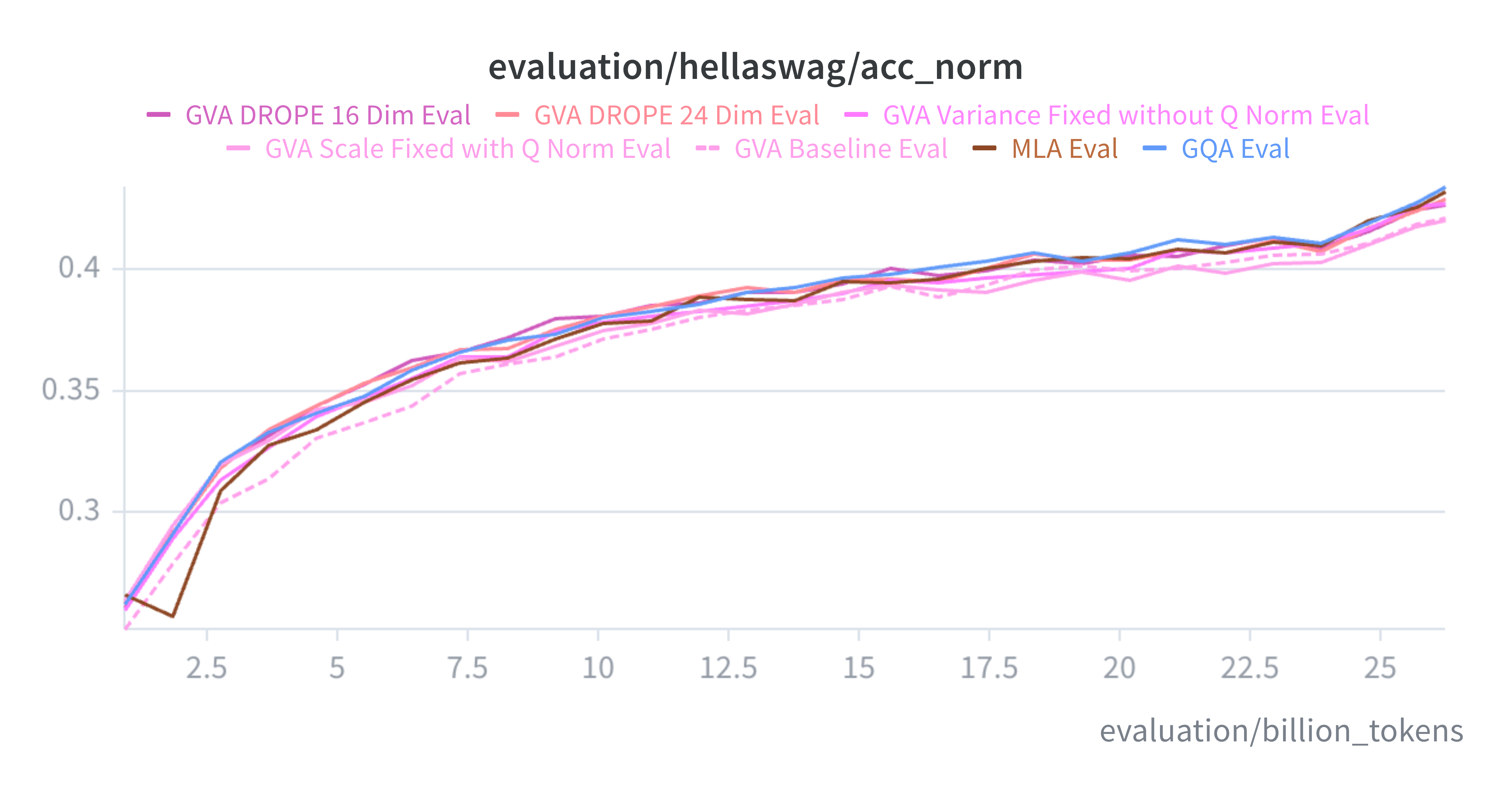}
\captionof{figure}{HellaSwag accuracy over training.}
\label{fig:eval-hs}
\end{minipage}

\vfill
\noindent\begin{minipage}{\textwidth}
\centering
\includegraphics[width=0.95\textwidth]{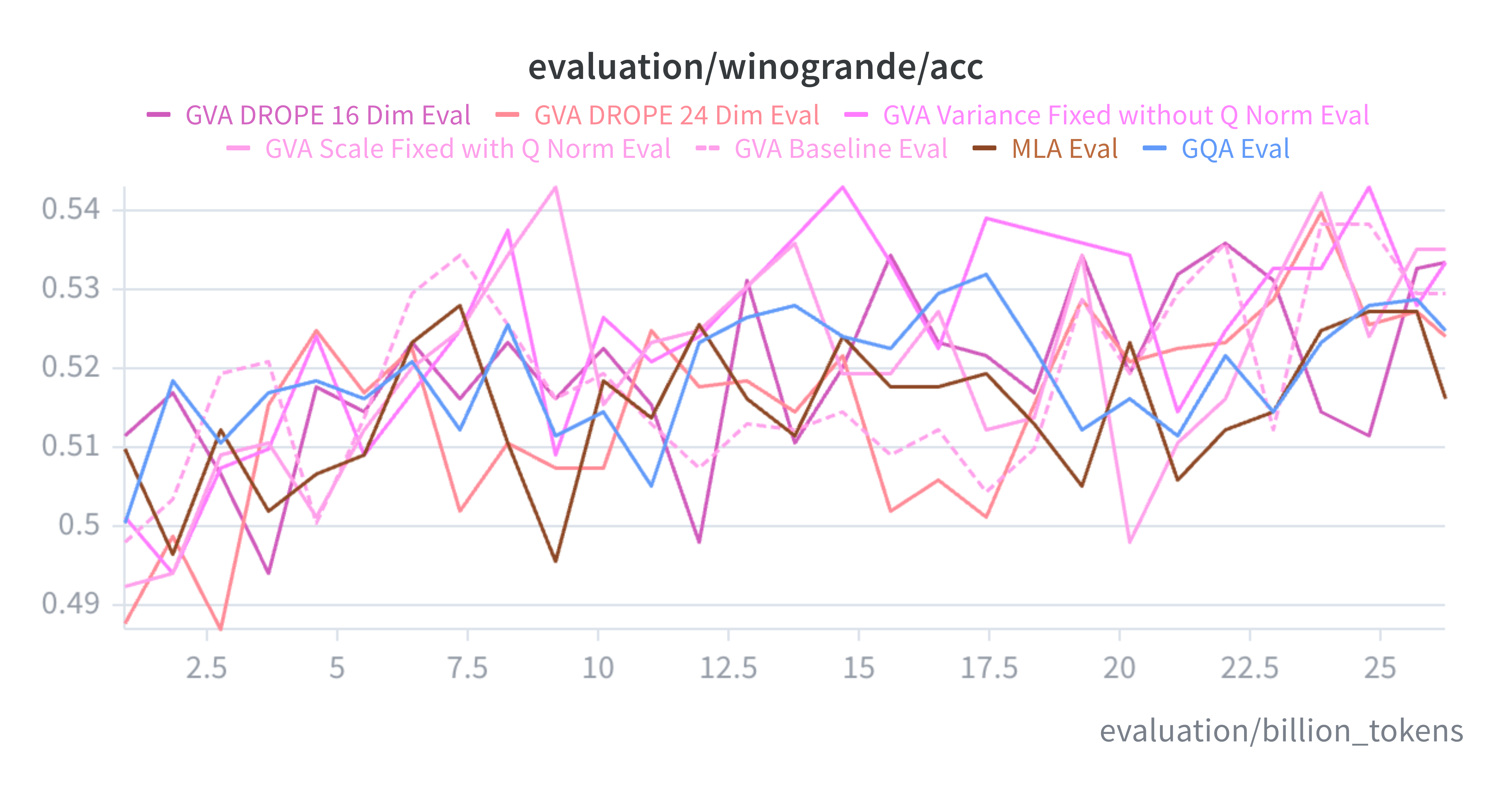}
\captionof{figure}{WinoGrande accuracy over training.}
\label{fig:eval-wg}
\end{minipage}
\clearpage

\noindent\begin{minipage}{\textwidth}
\centering
\includegraphics[width=0.95\textwidth]{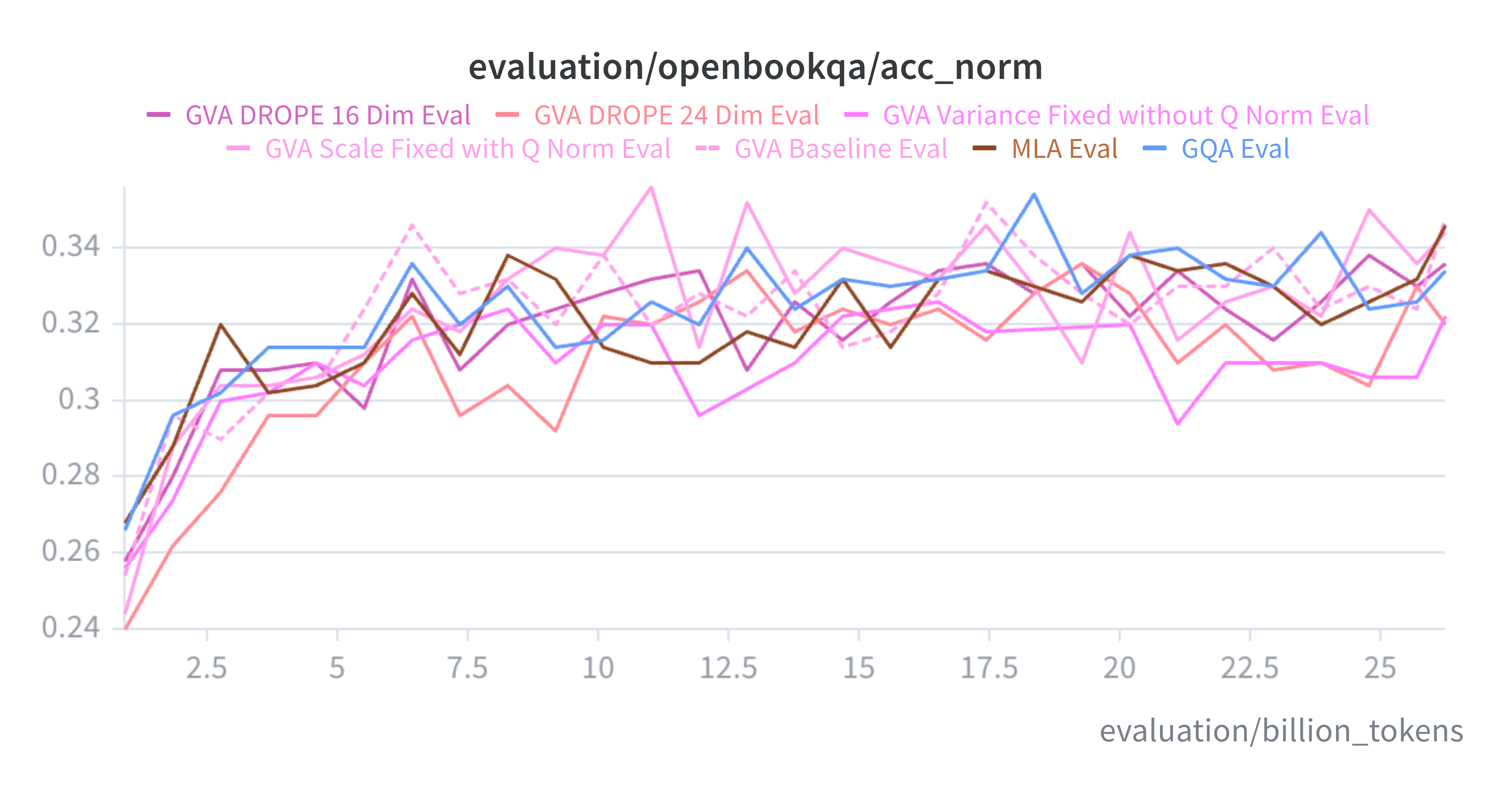}
\captionof{figure}{OpenBookQA accuracy over training.}
\label{fig:eval-obqa}
\end{minipage}

\vfill
\noindent\begin{minipage}{\textwidth}
\centering
\includegraphics[width=0.95\textwidth]{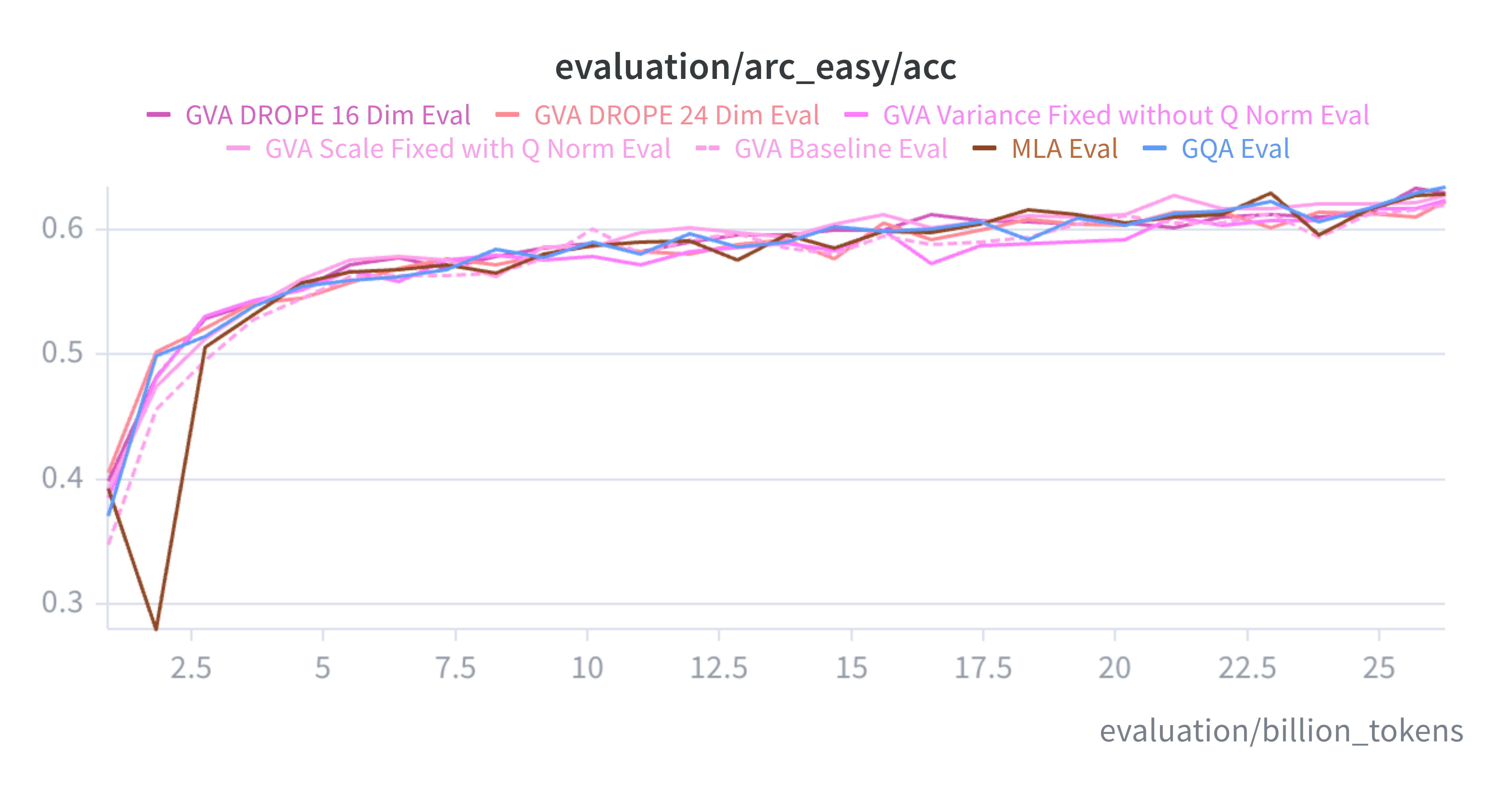}
\captionof{figure}{ARC-Easy accuracy over training.}
\label{fig:eval-arce}
\end{minipage}
\clearpage

\noindent\begin{minipage}{\textwidth}
\centering
\includegraphics[width=0.95\textwidth]{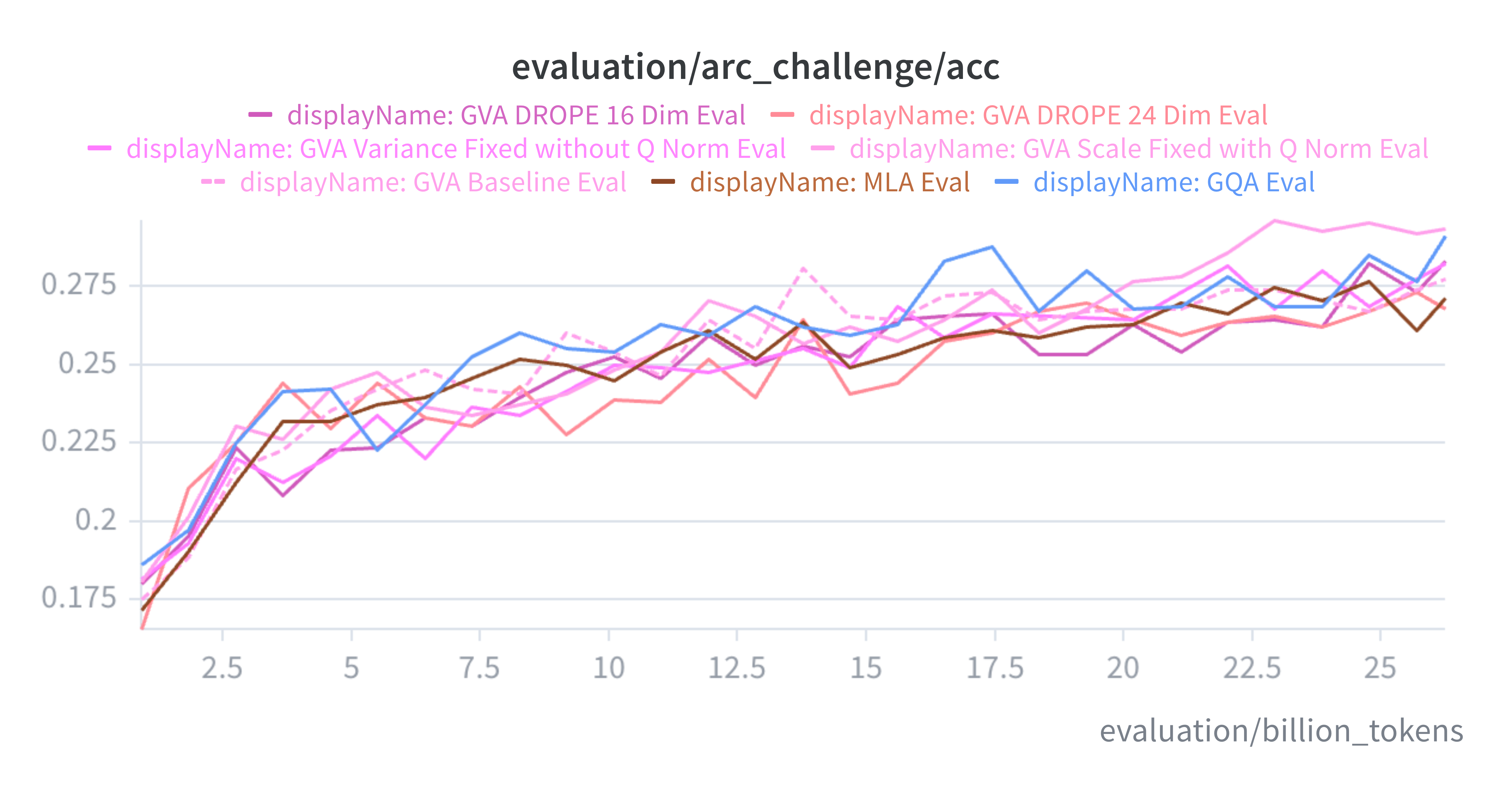}
\captionof{figure}{ARC-Challenge accuracy over training.}
\label{fig:eval-arcc}
\end{minipage}
\vfill
\clearpage
\endgroup

\end{document}